\documentclass{elsarticle}
\makeatletter
\def\ps@pprintTitle{%
	\let\@oddhead\@empty
	\let\@evenhead\@empty
	\let\@oddfoot\@empty
	\let\@evenfoot\@oddfoot
}
\makeatother
\usepackage{bm}

		\newcommand{\mnr}{{\mu \nu \rho }}

\usepackage{amssymb}

\begin{document}

\begin{frontmatter}



\title{Gauge theories for  fluids  in $2+1$ dimensions 
	through  master actions}

\author{\"{O}mer F. Dayi}

\address{\normalsize{Physics Engineering Department, Faculty of Science and Letters, Istanbul Technical University, Maslak, Istanbul, TR-34469, Turkey}}

\begin{abstract}
Two actions which are functionals  of different variables but  describing the same dynamical  system  can be shown to possess the same origin by constructing a master action which generates both of them.  We first  present the master action which produces an action  depending on the fluid variables and a gauge theory action  whose equations of motion are equivalent to the incompressible fluid Euler equations in $2+1$ dimensions.
We then introduce the  master action generating the actions  which on shell provide the linearized shallow water equations. One of them  is a functional  of
the variables of the shallow water  and the other one  is the  Maxwell-Chern-Simons gauge theory action.  
The maps between gauge vector fields and fluid variables are obtained for both of the systems. We employ them to derive the corresponding gauge theory solutions of the Hopfion  and the coastal  Kelvin wave solutions.

\end{abstract}


	
	
	



	
	
	

\end{frontmatter}


\section{Introduction}
A dynamical system can be  described through the differential equations of   its variables. When the solutions of these differential equations manifest some topological behaviour, an effective field theory formulation of the system  may shed light on  its topological properties.  In $2+1$ dimensions the most studied field theory in this respect is the Chern-Simons theory.   It recently featured  in \cite{tong1} where  an effective gauge field theory with a Chern-Simons term has been proposed to get an insight into the topological features of  the  shallow water  equations. 

Consider a thin layer of  rotating fluid whose bottom lies over a rigid plane but its  surface is not bounded. It is described by  the shallow water equations (see e.g. \cite{vallis})
\begin{eqnarray}
\label{sw1}	\partial_0 h+\bm u \cdot \bm \partial h +h \bm \partial \cdot \bm u  =  0, \\
\label{sw2}	\partial_0 u^i +\bm u \cdot \bm \partial  u^i  = f\epsilon^{ij} u_j -g \partial^i h .
\end{eqnarray}
The repeated indices are summed over. $h$ is the height of the fluid and  $u_i;\ i,j=1,2,$ denote  the components of the velocity vector $\bm u.$   They  are functions of the horizontal coordinates $x_i=(x,y)$ and time $t.$   $\partial_i\equiv \partial /\partial x^i,$ and $\partial_0 \equiv \partial /\partial t,$  indicate the derivatives, $\epsilon^{ij}$   is the antisymmetric symbol. $g$ is the acceleration due to gravity and $f$ is the constant Coriolis parameter.
In \cite{tong1} an Abelian $2+1$ dimensional gauge theory formulation of these equations is given.  To accomplish a consistent formulation,   a  Clebsch  (parametrized) decomposed auxiliary gauge field which is coupled to the $U(1)$ gauge theory has been introduced. 
There, $h$ and $u^i$ correspond, respectively to $B$ and $\epsilon^{ij}E_i/B ,$ where $B$ and $E_i$ are the magnetic and electric fields. 

The linearized shallow water equations  are obtained from (\ref{sw1}),(\ref{sw2}) by writing $h (x,y,t)=H + \eta (x,y,t)$ and ignoring the terms which are higher than first order  in $\eta$ and $\bm u:$  
\begin{eqnarray}
	\partial_0 \eta + H \bm \partial \cdot \bm u  = 0, \label{lin1} \\
	\label{lin2}
	\partial_0  u_i  = -g \partial_i    \eta +f  \epsilon_{ij}u^j .
\end{eqnarray}
(\ref{lin2}) leads to the conservation of  the vorticity  $\omega = \epsilon^{ij}\partial_i u_j:$ 
\begin{equation}
	\label{cvor}
	\partial_0  \omega +f \bm \partial\cdot \bm u =0.
\end{equation}  
By subtracting (\ref{lin1}) from  (\ref{cvor}) after  multiplying them, respectively, with $f$ and $H,$ one gets
the conservation of $Q\equiv H\omega -f\eta :$
\begin{equation}
	\partial_0 Q=0.
\end{equation}
In \cite{tong1} it is shown that for $Q=0$ the linearized rotating shallow water equations can be expressed as  the equations of motion following from the Maxwell-Chern-Simons (MCS) action.  Then, the edge modes of the MCS theory are associated with the  coastal Kelvin waves of shallow water.
Gauge symmetry features of this formalism is explored further  in \cite{sahin}. 

In \cite{ns} the compressible Euler fluid has been studied following the formulation of \cite{tong1}. There the   
fluid density  $\rho$ and the fluid velocity $\bm u$ correspond to the electromagnetic fields as
$B=\rho, $ $ E_i =\rho \epsilon_{ij} u^j.$  Like \cite{tong1}, they  couple  a Clebsch decomposed auxiliary gauge field  to the gauge fields.  
One of their goals was to uncover the  topological aspects of the  $2+1$ dimensional Hopfion solution of the incompressible Euler fluid  \cite{hopf1}, \cite{hopf2}.

An alternative  gauge field theory for  the  incompressible Euler fluid in $2+1$ dimensions has been constructed in \cite{eling}.   As we will discuss in detail, in this approach the magnetic field corresponds to  the vorticity. Although this formulation has also been  
inspired by \cite{tong1},  it is distinguished from that by not having need of introducing a Clebsch parametrized  auxiliary gauge field.

In \cite{tong1}  the gauge theory for shallow water was accomplished mainly by considering the conserved quantities and their continuity equations. However, there  it has also been demonstrated that one can achieve the same goal   by  considering  an action which is  a functional of the fluid variables, gauge fields and some Lagrange multiplier fields coupled to the $U(1)$ 3-vector potentials via a Chern-Simons like term. By replacing the fluid variables with the electromagnetic fields as implied by  the equations of motion derived by varying the action with respect to the Lagrange multipliers, one  gets  the desired gauge theory action. Instead, when one eliminates the Lagrange multiplier fields  by making use of the solutions of the equations of motions imposed  by the gauge fields, an action which produces the shallow water equations follows.  

The master actions which we present depend on the $U(1)$ 3-vector potential, fluid velocity and an auxiliary field which will be identified with the vorticity.
We deal with the incompressible fluid equations and the linearized shallow water equations  for $Q=0.$    In each case we will show that by getting rid of the gauge fields, the proposed  master action  generates an  action in fluid variables, $\bm u,$ whose equations of motion are  the related Euler equations. 
On the other hand, the associated gauge theory actions results if one eliminates the fluid variables in the master actions by imposing some of the equations of motions.
The master action formulation \cite{deser-jackiw} has a lot of advantages, for example it will permit us to directly construct the field theory configurations corresponding to the solutions of the Euler equations.

In Section \ref{IFE}  the master action for the incompressible fluid will be presented.  We will demonstrate that it generates either the action depending on the fluid variables whose equations motion are  the incompressible Euler fluid equations or the associated gauge theory action within the formulation of \cite{eling}. 
 We will present the gauge theory solution which correspond to the $2+1$ dimensional Hopfion solution of the fluid. 
In Section  \ref{SW} we will show that a similar approach is available for the  linearized shallow water equations for $Q=0.$ 
We will introduce the  master action and derive the  daughter actions. One of them is a functional of fluid variables and the other one is the MCS action. Their equations of motion are equivalent to the linearized equations of shallow water. We will  present the map between the coastal Kelvin wave solution  and the 3-vector potential.  In Section \ref{dis} we  discuss our results.

\section{Incompressible Fluid}
\label{IFE}

By suppressing the constant  density,
the incompressible fluid Euler equations in $2+1$ dimensions are given as
\begin{eqnarray}
	\bm \partial \cdot \bm u& = & 0,  \label{e1} \\
	\partial_0 u_i + \bm u   \cdot \bm \partial  u_i +\partial_i P & = & 0, \label{e2}
\end{eqnarray}
where $P$ is the fluid pressure which does not depend  explicitly on  time.  One can easily show that the vorticity  satisfies 
\begin{equation}
	\partial_0  \omega +\bm u   \cdot \bm \partial\,  \omega =0.
\end{equation}  
We solve (\ref{e1})  in terms the stream function $\psi$ as
\begin{equation}
	u_i =\epsilon_{ij} \partial^j \psi.
\end{equation}
Therefore, we only deal with the Euler equation (\ref{e2}).   
To associate a gauge theory to the Euler fluid one needs to be acquainted with the explicit form of the pressure $P.$ 
For our purposes it is sufficient that it can be written in terms of the fluid velocity and its spatial derivatives.  We pick it out as 
\begin{equation}
	\label{pres}
	P=P_0+ P (\omega)\equiv  P_0 +\kappa \omega^n,
\end{equation}
where $P_0$ is a constant. The  constant   $\kappa$  and the  real number $n$  will be dictated by the given pressure
as it will be clarified shortly. 
However, we can anticipate that (\ref{pres}) is chosen to be adequate for the Hopfion solution.

We propose the following master action
\begin{equation}
	\label{mf}
	S_{mf} =	\int d^3x \left[\frac{1}{2} \theta u^2 -\theta A_0 -\theta u^i A_i  
	-\frac{\kappa\, \theta^{n+1} }{n+1}+\frac{1}{2} \epsilon^{\mnr} A_\mu \partial_\nu A_\rho \right] ,
\end{equation}
where $\theta, u_i$ and $A_\mu =(A_0, A_1,A_2)$ are independent fields. 
We will derive first the gauge theory for the incompressible Euler fluid. We will then show that the master action (\ref{mf}) generates
a fluid theory action which  yields the Euler equation (\ref{e2}).

\subsection{Gauge Theory Action }

By varying the master action (\ref{mf})  with respect to $A_\mu ,$ we acquire the equations of motion
\begin{eqnarray}
	\theta &=& 	B , \label{bt}\\
	\theta \epsilon_{ij}u^j &= &  E_i  ,	\label{et}
\end{eqnarray}
where $B=\epsilon^{ij} \partial_i A_j$ and $E_i=\partial_0 A_i -\partial_i A_0.$
By replacing $\theta$ and $u_i$ with the gauge fields as dictated by (\ref{bt}) and (\ref{et}) in  (\ref{mf}),  we arrive at
\begin{equation}
	\label{sF}
	S_{F} =	\int d^3x \left[\frac{E^2}{2B} 
	-\frac{\kappa\, B^{n+1} }{n+1}- \frac{1}{2} \epsilon^{\mnr} A_\mu \partial_\nu A_\rho \right] .
\end{equation}
The equations of motion  following from the gauge theory action (\ref{sF})  are
\begin{eqnarray}
	\label{emao}
	\partial_i \left( \frac{E^i}{B}\right) -B &=& 0,\\
	\label{gte}
	-\partial_0  \left( \frac{E^i}{B}\right) -\epsilon^{ij} \partial_j\left( \frac{E^2}{2B^2}\right) -\epsilon^{ij} \partial_jP(B)+ \epsilon^{ij} E_j  &= &0.
\end{eqnarray}
By inserting (\ref{bt}) into  (\ref{et})  one gets
\begin{equation}
	\label{tom1}
	\frac{E_i}{B} = \epsilon_{ij}u^j .
\end{equation}
Then,  by making use of (\ref{tom1})  in   (\ref{emao}),  we deduce that actually the magnetic field  corresponds to the fluid vorticity: 
\begin{equation}
	\label{tom}
	B =\omega.
\end{equation}
By plugging (\ref{tom1}) and  (\ref{tom}) into the equation of motion  (\ref{gte}), one gets
\begin{equation}
	\label{ee2}
	\partial_0  u_i +u_j \partial_i u^j -\omega \epsilon_{ij}u^j + \partial_i P (\omega)=0.
\end{equation}
Now, by using  the identity
$$
\partial_i u_j =\epsilon_{ij} \omega +\partial_j u_i,
$$
one can easily observe that (\ref{ee2}) is the same with  the Euler equation (\ref{e2}). 

The variation of (\ref{sF}) with respect to $A_\mu$ leads to the equations of motion  (\ref{emao})-(\ref{gte}),  as far as  the  boundary terms vanish.   If there is a  boundary  at $x_i=0,$ the boundary terms can be shown to be
\begin{equation}
	\label{bcf}
	\delta S_F =
	\int dt [d\hat{x}]_i \left[\left(-\frac{E^i}{B} +\frac{1}{2}\epsilon^{ij}A_j\right)\delta A_0 -\left( \frac{E^2}{2B^2} +P(B) +\frac{1}{2}A_0\right) \epsilon^{ij}\delta A_j
	\right],
\end{equation}
where $[d\hat{x}]_i=(dx_2,dx_1).$ Thus,  the condition  $\delta S_F =0,$ should also be fulfilled by  the solutions of (\ref{emao})-(\ref{gte}).

\subsection{Fluid action }
By varying (\ref{mf}) with respect to the fields $u_i$ and $\theta$ we get
\begin{eqnarray}
	\theta \left(u_i -A_i\right) &=& 0, \label{aii} \\
	\frac{1}{2} u^2-A_0-u^iA_i -P (\theta) &=& 0. \label{a00} 
\end{eqnarray}
As far as  $\theta \neq 0,$  one can solve   (\ref{aii}) and (\ref{a00})  as
\begin{eqnarray}
	A_i &=& u_i \label{m1} \\
	A_0 &=	&-\frac{1}{2} u^2- P (\theta) .\label{m2} 
\end{eqnarray}
By plugging (\ref{m1}), (\ref{m2})  into (\ref{mf}) and  integrating by parts we acquire
\begin{equation}
	\label{f}
	S_{1} =	\int d^3x \left[-\frac{1}{2}\epsilon^{ij}u_i \partial_0 u_j -\frac{1}{4} \omega u^2  -\frac{1}{4} \epsilon^{ij} u_i\partial_j u^2	
	+\frac{n\kappa}{n+1}\theta^{n+1} -\kappa \omega \theta^{n} \right] .
\end{equation}
It still depends on the auxiliary field $\theta$  due to the pressure. To express  it  in terms of the fluid variables,
we employ the equation of motion derived by varying  (\ref{f}) with respect to  $\theta :$ 
$$
\kappa n \theta^{n-1} (\theta -\omega)=0,
$$
which is solved as
$$
\theta =\omega.
$$
Hence the action in terms of the velocity and vorticity follows:
\begin{equation}
	\label{L}
	S_{IF}=	\int d^3x {\cal L} =	\int d^3x\left[-\frac{1}{2}\epsilon^{ij}u_i \partial_0 u_j -\frac{1}{4} \omega u^2  -\frac{1}{4} \epsilon^{ij} u_i\partial_j u^2	
	- \frac{\kappa}{n+1}\omega^{n+1}  \right] .
\end{equation}
One can show that 
the equations of motion
$$
\frac{\partial  {\cal L} }{\partial u_i} -\partial_\nu \left[\frac{\partial  {\cal L} }{\partial (\partial_\nu u_i)}\right] =0,
$$
lead to  (\ref{ee2}). Hence, we conclude that (\ref{L}) generates the incompressible fluid Euler equation (\ref{e2}).

\subsection{Hopfion solution}
The $2+1$ dimensional Hopfion  solution  is given  in \cite{hopf1}, \cite{hopf2} by
\begin{eqnarray}
	u_i& = & \frac{2\epsilon_{ij}x^j}{1+x^2+y^2} \, ,\\
	P & = & P_0 -\frac{2}{1+x^2+y^2}  \cdot
\end{eqnarray}
The vorticity can be calculated as
\begin{equation}
	\omega =-\frac{4}{(1+x^2+y^2)^2} \cdot
\end{equation}
Then, one observes that the pressure is a function of the vorticity:
\begin{equation}
	P =P_0- \sqrt{|\omega |}.
\end{equation}
This justifies the choice of the pressure as in
(\ref{pres}), where now $\kappa =-1$ and $n=1/2.$
We can read the corresponding gauge theory solutions from (\ref{m1}) and (\ref{m2}) as
\begin{eqnarray}
	A_i &=&  \frac{2\epsilon_{ij}x^j}{1+x^2+y^2} , \label{m11} \\
	A_0 &=& 	\frac{2}{(1+x^2+y^2)^2} \cdot \label{m22} 
\end{eqnarray}
Observe that they satisfy the gauge condition $\partial_\mu A^\mu=0.$
Electric and magnetic fields are 
\begin{eqnarray}
	E_i &=&  \frac{8x_i}{(1+x^2+y^2)^3}, \label{E1} \\
	B &=& 	-\frac{4}{(1+x^2+y^2)^2}.\label{B2} 
\end{eqnarray}
One can easily verify that the equations of motion (\ref{gte}) are satisfied by (\ref{E1}) and (\ref{B2}).

Let there be  a boundary at $x=0.$ In this case the boundary terms  (\ref{bcf}) are
\begin{equation}
	\label{bcfx}
	\delta S_F =\int dt dy  \left[\left(-\frac{E_1}{B} +\frac{1}{2}A_2\right)\delta A_0 -\left( \frac{E^2}{2B^2}  +P(B)+ \frac{1}{2}A_0\right) \delta A_2.
	\right]_{x=0}.
\end{equation}
For the Hopfion solution  
$$
-\frac{E_1}{B} \Bigg |_{x=0}=0,\ \ \ \ \ \   A_2 \big |_{x=0}=0.
$$
Therefore, the boundary terms in (\ref{bcfx}) vanish. Obviously a boundary  at $y=0$   can be considered in the same manner.

\section{Shallow Water }
\label{SW}

We deal with the linearized shallow water equations for $Q=H\omega -f\eta =0$ \cite{tong1}. The master action  which we propose is 
\begin{equation}
	\label{msw}
	S_{msw} =	\int d^3x \left[\frac{H}{2} u^2 -\frac{gH^2}{2f^2}\theta^2-\theta A_0 -f u^i A_i +\frac{f}{2H} \epsilon^{\mnr} A_\mu \partial_\nu A_\rho \right] .
\end{equation}
As before  $\theta, u_i$ and $A_\mu $ are independent field variables.  We will first show that (\ref{msw}) generates the MCS action. Then we
will demonstrate  that it also produces an action whose equations of motion are actually  the linearized shallow water equations for $Q=0.$ 

\subsection{Gauge theory action}

The variation of (\ref{msw})  with respect to  $A_\mu$ yields
\begin{eqnarray}
	\theta &=& 	\frac{f}{H}B , \label{bts}\\
	H \epsilon_{ij}u^j &= &  E_i  . 	\label{ets}
\end{eqnarray}
By replacing $\theta$ and $u_i$ with the electric and magnetic fields as in (\ref{bts})-(\ref{ets}),  the master action (\ref{msw}) gives rise to  the MCS  action
\begin{equation}
	\label{MCS}
	S_{MCS} =	\int d^3x   { \frac{1}{2H} } \left[E^2 - c^2B^2- f \epsilon^{\mnr} A_\mu \partial_\nu A_\rho \right] 
\end{equation}
where $c^2=gH.$ Its  equations of motion are
\begin{eqnarray}
	\label{emaos}
\bm	\partial \cdot \bm  E  =fB, \\
	\label{gtes}
	\partial_0  E_i = -c^2\epsilon^{ij} \partial_j B +f\epsilon^{ij}  E_j .
\end{eqnarray}
By making use of (\ref{bts}) and  (\ref{ets}) in  (\ref{emaos})  we notice that the magnetic field corresponds to the vorticity:
\begin{equation}
	\label{toms}
	B =\frac{H}{f} \omega.
\end{equation}
Then, by inserting (\ref{ets}) and  (\ref{toms}) into the equations of motion  (\ref{gtes}) we get
\begin{equation}
	\label{ee2s}
	\partial_0  u_i = -\frac{c^2}{f} \partial_i    \omega +f  \epsilon_{ij}u^j .
\end{equation}
Because of dealing with $Q=0,$  (\ref{ee2s})  is equal to the linearized shallow water equation (\ref{lin2}). Recall that (\ref{lin1}) has already been taken into account in the derivation of  $\partial_0 Q=0.$ 

In the presence of  a  boundary  at $x_i=0,$ the variation of (\ref{MCS}) furnishes the equations of motion (\ref{emaos})-(\ref{gtes}),  as far as the  the following  boundary terms vanish,
\begin{equation}
	\label{mcsd}
	\delta S_{MCS} =
	\int dt [d\hat{x}]_i  \left[\left(-2E^i+f\epsilon^{ij}A_j\right)\delta A_0 -\left( 2c^2B +fA_0\right) \epsilon^{ij}\delta A_j
	\right] .
\end{equation}
Recall that  $[d\hat{x}]_i =(dx_2,dx_1).$

\subsection{Fluid Action}

By  varying   the master action  (\ref{msw})  with respect to the fields $u_i$ and $\theta,$ we acquire
\begin{eqnarray}
	Hu_i &=&fA_i  , \label{aai} \\
	-\frac{gH^2}{f^2} \theta & = & A_0. \label{aa0}
\end{eqnarray}  
Now, by replacing the gauge fields with the fluid variables as given in (\ref{aai}) and (\ref{aa0}), one can show that
up to an integration by parts, (\ref{msw}) yields
\begin{equation}
	\label{SS2}
	S_{S2}= \int d^3x H \left[ -\frac{1}{2f}\epsilon_{ij} u^i \partial_0  u_j  -\frac{1}{2}u^2 +\frac{gH}{2f^2}\theta^2- \frac{gH}{f^2}\theta \omega  \right].
\end{equation}
The equation of motion following from the variation of (\ref{SS2})  with respect to $\theta$ is
$$
\theta=\omega.
$$
By substituting the vorticity $\omega$ for $\theta$  in (\ref{SS2}), one  acquires the action
\begin{equation}
	S_S= \int d^3x {\cal L}_S= \int d^3x  H \left[ -\frac{1}{2f}\epsilon_{ij} u^i \partial_0  u_j  -\frac{1}{2}u^2 -\frac{c^2}{2f^2}\omega^2  \right].
\end{equation}
One can show  that the equations of motion
$$
\frac{\partial  {\cal L}_S }{\partial u_i} -\partial_\nu \left[\frac{\partial  {\cal L_S} }{\partial (\partial_\nu u_i)}\right] =0,
$$
lead to  (\ref{ee2s}).  Therefore, they are equivalent to the linearized shallow water equation  (\ref{lin2}).

We conclude that the map between the vector fields and the fluid variables is given by
\begin{equation}
	A_0 =-\frac{c^2H}{f^2}\omega ,\ \ \ 
	A_i = \frac{H}{f} u_i \label{aiu} .
\end{equation}  
Recall that solutions of the MCS action should also satisfy the boundary condition $	\delta S_{MCS} =0.$

\subsection{Kelvin waves}

For a solid boundary  at $y=0$ Kelvin waves are described (see e.g. \cite{vallis}) by
\begin{eqnarray}
	u_1 &=& e^{-fy/c} G (x-ct), \\
	u_2 &=&0, \\
	\eta &=&\frac{H}{c} e^{-fy/c} G (x-ct).
\end{eqnarray}
$ G (x-ct)$ is an arbitrary function. Now the map (\ref{aiu})
yields
\begin{eqnarray}
	A_0 & =& \frac{cH}{f} e^{-fy/c} G (x-ct) , \label{aok} \\
	A_1 & =& -\frac{H}{f} e^{-fy/c} G (x-ct) , \label{a1k} \\
	A_2 &= &0. \label{a2k} 
\end{eqnarray}
Then, the electric and magnetic fields are calculated  as 
\begin{eqnarray}
	E_1 & =& 0, \label{e1k} \\
	E_2 & =& -H e^{-fy/c} G (x-ct), \label{e2k} \\
	B &= & -\frac{H}{c} e^{-fy/c} G (x-ct) . \label{bk} 
\end{eqnarray}
One can observe that the field equations (\ref{emaos}) and  (\ref{toms}) are satisfied. Moreover, we can choose
$$
G (x-ct) = A \exp(w t-kx) 
$$
where $A$ denotes a constant amplitude, $k$ is the wave number and  $w=kc.$ 

For a solid boundary at $y=0,$  the equations  of motion should be supplemented with  the boundary  terms
$$
\delta S_{MCS} =\int dt\, dx \left[-\left(2E_2 +fA_1\right)\delta A_0+ \left(2c^2B +fA_0\right)\delta A_1\right] .
$$
Obviously,  it vanishes if on the boundary $\delta A_0=c\delta A_1.$ In fact, for the Kelvin waves (\ref{aok}), (\ref{a1k}), $A_0=cA_1$ everywhere.

\section{Discussions}
\label{dis}

To explore the basic properties of theories which are dual to each other it is  useful to be acquainted with the related master action \cite{deser-jackiw}, which generates  the actions of the dual theories.  Here, we introduced the master actions  to clarify the duality between the incompressible fluid and shallow water equations with the associated  gauge theories. 

We first discussed, following basically the approach of \cite{eling},  the correspondence between the incompressible Euler fluid fields and the electromagnetic fields.  We  presented the master action which generates the Lagrange density whose equations of motion coincide with the Euler equations  and  the action of the associated gauge theory. We presented  the map between the Hopfion solutions of the fluid and  the  3-vector gauge  fields. Then, we proposed  a master action for studying the duality between  the linearized   shallow water equations and the MCS gauge theory.  We explicitly obtained the gauge fields corresponding to the coastal Kelvin waves of the linearized rotating shallow water equations.

The main motivation of  studying these dual theories is to have a better understanding of the topological aspects of fluids.  In this respect the explicit map which we obtained between the fluid variables and vector fields of the underlying gauge theory which possesses topological sectors, would be  helpful. 
We discussed the incompressible fluid equations,  it  would be nice to extend our discussion to the compressible fluid equations  and obviously to the fluid equations with the fluid  viscosity term.




\bibliography{fluid-gauge-els}
\end{document}